\documentclass[manuscript]{aastex63}

\usepackage{lineno}

\newcommand{\be}{\begin{equation}} 
\newcommand{\ee}{\end{equation}} 
\newcommand{\bea}{\begin{eqnarray}} 
\newcommand{\eea}{\end{eqnarray}} 

\received{Feb 27, 2021}
\revised{Mar 8, 2021}
\accepted{Mar 9, 2021}
\submitjournal{ApJL}

\shorttitle{Asteroid Gault}
\shortauthors{Luu et al.}

\begin{document}

\title{Rotational Mass Shedding from Asteroid (6478) Gault}


\correspondingauthor{Jane X. Luu}
\email{janeluu@uio.edu}

%

\author{Jane X. Luu}
\affiliation{Centre for Earth Evolution and Dynamics, Department of Geosciences, University of Oslo}
\affiliation{Institute of Theoretical Astrophysics, University of Oslo \\
P.O. Box 1047, Blindern, NO-0316 Oslo, Norway}

\author{David C. Jewitt}
\affiliation{Department of Earth, Planetary and Space Sciences, UCLA \\
595 Charles Young Drive East, Los Angeles, CA 90095-1547}
\affiliation{Department of Physics and Astronomy, UCLA \\
430 Portola Plaza, Box 951547, Los Angeles, CA 90095-1567}

\author{Max Mutchler}
\affiliation{Space Telescope Science Institute \\
3700 San Martin Drive, Baltimore, MD 21218}

\author{Jessica Agarwal}
\affiliation{Institut for Geophysik und Extraterrestrische Physik, Technische Universitat Braunschweig \\
Mendelssohnstr. 3, 38106 Braunschweig, Germany}

\author{Yoonyoung Kim}
\affiliation{Institut for Geophysik und Extraterrestrische Physik, Technische Universitat Braunschweig \\
Mendelssohnstr. 3, 38106 Braunschweig, Germany}

\author{Jing Li}
\affiliation{Department of Earth, Planetary and Space Sciences, UCLA \\
430 Portola Plaza, Box 951547, Los Angeles, CA 90095-1567}

\author{Harold Weaver}
\affiliation{The Johns Hopkins University Applied Physics Laboratory \\
1100 Johns Hopkins Road, Laurel, Maryland 20723}


%

%
%
%
%
%



\begin{abstract}

The $\sim$4 km diameter main belt asteroid 6478 Gault has ejected dust intermittently since at least 2013.  The character of the emission, including its episodic nature and the low  speed of the ejected particles ($V \sim $ 0.15 m s$^{-1}$), is most consistent with mass loss from a body rotating near rotational breakup. Owing to dust contamination of the nucleus signal, this conclusion has not yet been confirmed. 
To test this idea, we have obtained new images of Gault in August 2020, in the absence of dust.  Our photometry shows a lightcurve having a very small amplitude (maximum $\sim 0.05$ mag) and a periodicity of $ 2.55 \pm 0.10$ hours.  The new observations are consistent with a model in which Gault is rotating near breakup, with centrifugal forces responsible for its episodic mass loss.  Approximated as a strengthless (fluid) spherical body, the implied density is $\rho$ = 1700 kg m$^{-3}$.  We use the Froude number $Fr$, defined here as the ratio between centrifugal force and gravitational force, as a way to investigate mass loss regimes in fast spinning asteroids and find that mass shedding starts at $Fr \sim 0.5$.

\end{abstract}



\section{Introduction} \label{sec:intro}

Main belt asteroid (6478) Gault (hereafter ``Gault'') was first reported to be active on 2019 January 8 (\cite{smith2019}). Several groups monitored the development of the activity, finding two (\cite{hui2019}, \cite{kleyna2019}, \cite{moreno2019}, \cite{ye2019}) and three (\cite{jewitt2019a}) tails, each corresponding to a discrete dust emission event.  Archival images additionally showed that Gault has been intermittently active since 2013 (\cite{chandler2019}), apparently at intervals unrelated to its perihelion.  The  dust grains ejected by Gault are large and slow moving, with radii in the $10^{-5} - 10^{-3}$ m range and speeds $V \sim 0.15$ m s$^{-1}$ (\cite{jewitt2019a}, \cite{kleyna2019}).  Variable optical colors, ranging from slightly blue to slightly red, may indicate compositional or particle size changes due to fallback of fresh dust onto the asteroid surface (\cite{marsset2019}, \cite{carbognani2020}).  \\

The orbit of Gault has semimajor axis $a$ =  2.306 AU, eccentricity $e$ = 0.193 and inclination $i$ = 22.8\degr.  The resulting Tisserand parameter with respect to Jupiter is $T_J$ = 3.46.  This combination of asteroidal orbit and comet-like mass loss together establish  Gault as a member of the  Active Asteroids population (\cite{jewitt2015}), currently numbering about two dozen objects.   Although dynamically distinct from comets, some active asteroids  exhibit repeated mass loss approximately in phase with perihelion, while others (like Gault) show no relation to perihelion or are recorded as a single occurrence only.  The periodic, perihelion-active asteroids are best explained by the sublimation of near-surface ice, but other objects reflect a diverse range of physical processes from rotational instability, to thermal fracture, to impacts. In many cases the scarcity of data prevents pinpointing the responsible mechanism.  \\

Gault's irregular but repeated episodes of mass loss spread over many years are inconsistent with an origin by impact ejection which would be, presumably, impulsive and singular.  Its multiple short-duration dust releases, uncorrelated with perihelion, also distinguish it from those objects in which the activity is driven by water ice sublimation (\cite{jewitt2019a}).  By elimination, Jewitt {\it et al.} concluded that rotational instability was the most likely  cause of mass loss in Gault.  This hypothesis, however, could not be confirmed due to the presence of an extensive near-nucleus dust coma, which prevented the detection of a rotational lightcurve in their data.  A similar conclusion was also reached by \cite{moreno2019} and \cite{lin2020}.  On the other hand, \cite{kleyna2019} used high resolution observations from the Hubble Space Telescope to find a  period near $P \sim$ 2 hr, consistent with a rotational instability origin.   \cite{ferrin2019} reported a longer period, $P$ = 3.360$\pm$0.005,  while \cite{carbognani2020} found an apparently consistent value $P$ = 3.34$\pm$0.02 hr.  However, given the presence of substantial coma in the observations, these periods are all open to question.  \\

In this paper we present new observations of Gault taken in the absence of measurable near-nucleus dust. We use them to more reliably determine the nucleus rotation period, the most basic test of the rotational instability hypothesis. \\

\section{Observations} \label{sec:observations}

We obtained images of Gault  on UT 2020 August 26/27 with the 2.5m diameter Nordic Optical Telescope (NOT), located on La Palma, the Canary Islands.  The instrument used was the Andalucia Faint Object Spectrograph and Camera (ALFOSC) optical camera, equipped with an e2v Technologies charge-coupled device (CCD) with $2048 \times 2064$ pixels.  The camera had pixel scale 0.214"/ pixel, resulting in a vignette-limited field of view approximately $6.5 \times 6.5$ arcmin.  All observations were made in the broadband Bessel R filter (central wavelength $\lambda_c = 6500$ \AA ,  full-width at half max (FWHM) 1300  \AA).  We observed Gault almost continuously over a timespan of $\sim 5$ hours, with images each of 200 s integration. The image processing steps were as follows: the images were first bias subtracted and then normalized by a flat-field image constructed from images of the sky.  The telescope was tracked at Gault's angular rates ($\sim$10\arcsec/hour West and 30\arcsec/hour South) so  Gault had a stellar appearance while field stars were slightly trailed.  The observing conditions suffered from an unusually high level of Saharan dust in the atmosphere, resulting in a larger than normal extinction ($\sim 1$ magnitude / airmass).   This extinction was not uniform over the entire sky, rendering observations of Landolt stars unhelpful.  However, we established that the extinction was uniform over the CCD field of view, so we were able to perform differential photometry of Gault using nearby field stars.  The final photometric calibration of the data was obtained by comparing the stellar magnitude measured at zenith with photometry from the Sloan Digital Survey SkyServer DR14, using the ugriz to UBVRI photometric transformation from \cite{jester2005} \footnote{https://classic.sdss.org/dr7/algorithms/sdssUBVRITransform.html}.   Seeing was approximately 0.9" FWHM at low airmass, but increased to $\sim 1.2$ arcsec at large airmass. \\

\begin{figure}
\plotone{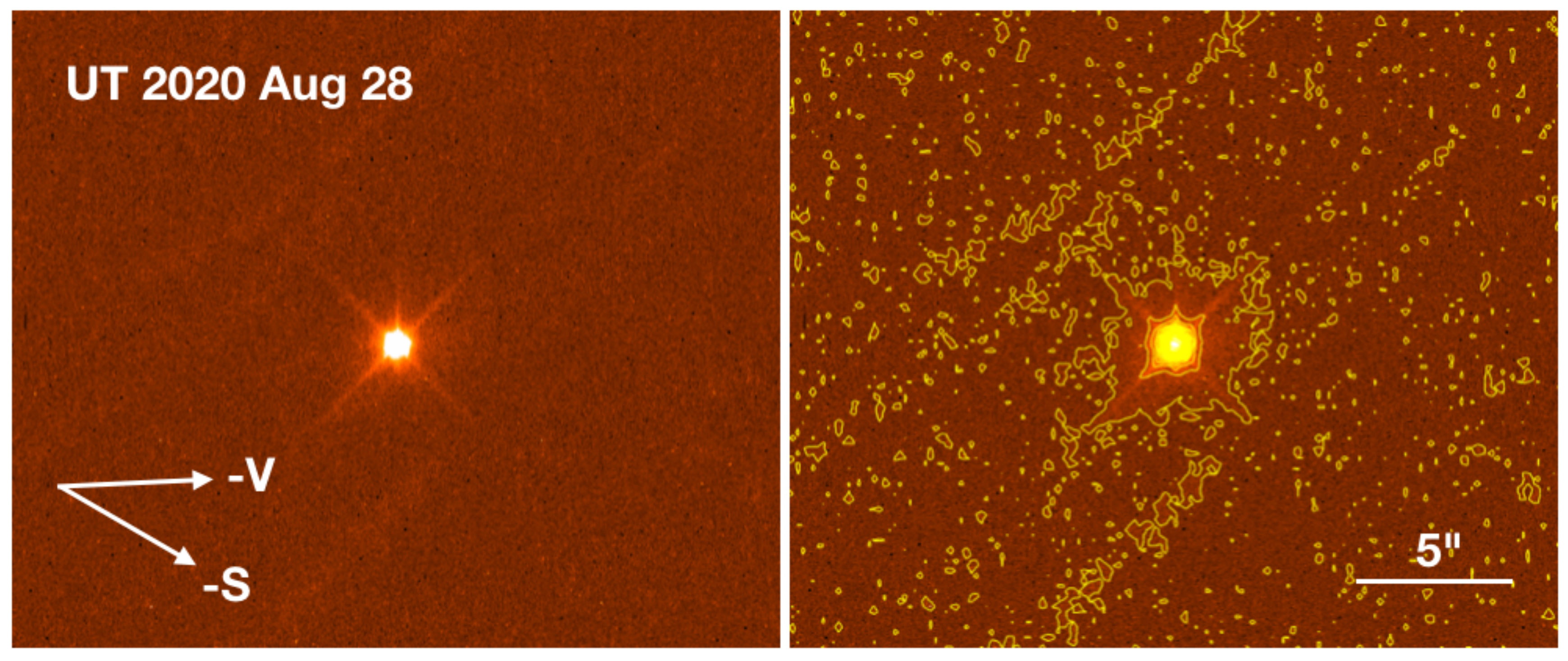}
\caption{Gault as observed by Hubble Space Telescope on UT 2020 August 28.  The image shows the median of 4 images.   -S = projected antisolar direction, -V = projected negative heliocentric velocity vector.  Faint residual structures are due to imperfect removal of field stars.  The field width is 24\arcsec.
\label{fig:HSTfig}}
\end{figure}

We also obtained four high resolution images of Gault with the Hubble Space Telescope (HST; GO 15972, PI Jewitt) to search the near-nucleus space for evidence of coma.  Taken on UT 2020 Aug 28, these images had 245 s exposure each, and were acquired with the WFC3 camera and the F350 LP broadband filter (central wavelength $6230 \AA$, FWHM = $4758 \AA$). The pixel size was 0.04\arcsec/pixel giving Nyquist resolution 0.08\arcsec.  The composite HST image is shown in Figure \ref{fig:HSTfig}, where faint trails result from imperfect removal of background objects.  The cross pattern result from diffraction around the secondary support arms while the vertical spike above the core is caused by imperfect charge transfer in the detector.  No coma is evident.  Observational parameters for  both NOT and HST observations are listed in Table 1.  \\


%
%

\subsection{Photometry}

Gault appeared stellar in all our images.  We measured the apparent magnitude  in the NOT images using circular apertures with angular radii 5, 10, and 15 pixels (1.1\arcsec, 2.1\arcsec, and 3.2\arcsec, respectively).  Sky subtraction used an annulus having inner radius 15 pixels (3.2\arcsec) and outer radius 50 pixels (10.8\arcsec).  The same apertures were used for field stars in order to obtain relative photometry.  The apparent magnitudes were then converted to Johnson R magnitude (\cite{jester2005}) and are plotted in Figure \ref{fig:lightcurve}. The data have a mean R magnitude $\overline{m}_R = 17.64$, with rms error $\pm 0.01$ mag.  To compare with previous works, we convert the mean R magnitude to V, adopting $m_V$ - $m_R$ =   0.40$\pm$0.01 (\cite{jewitt2019a}).  The  mean V magnitude is thus $\overline{m}_V = 18.04 \pm 0.02$, and is related to the mean absolute magnitude, $H$, by

\begin{equation}
H = m_V - 2.5\rm{log}_{10}\left(r_H^2 \Delta^2\right) + 2.5 log_{10} \left(\Phi \left( \alpha \right) \right)
\end{equation}

\noindent where $r_H$ and $\Delta$ are the heliocentric and geocentric distances expressed in astronomical unit (au) and $\Phi \left( \alpha \right)$  is the phase function measured at phase angle $\alpha$.  The absolute magnitude $H_V$  (Equation (1)) is the magnitude Gault would have if observed from $r_H =  \Delta = 1$ au and $\alpha = 0 ^{\circ}$.  We assume $2.5 \rm{log}_{10} \left( \Phi \left( \alpha \right) \right) = -0.04 \alpha$, broadly consistent with the measured phase function of asteroids.  Equation (1) yields $H = 15.0$, 0.7 magnitudes fainter than the $H = 14.3$  absolute magnitude listed in the JPL Horizons software catalog.  \cite{marsset2019} presents a summary of available photometry of Gault taken between January and May 2019 in their Figure 2.  This figure shows that the faintest absolute magnitude recorded during that period was $H = 14.8$, obtained at the end of April 2019.  Our NOT images, obtained more than a year later, show Gault in an even fainter state,  giving us confidence that we were observing the bare nucleus of Gault.   \\

\subsection{Surface Brightness Profile}

\begin{figure}
\plotone{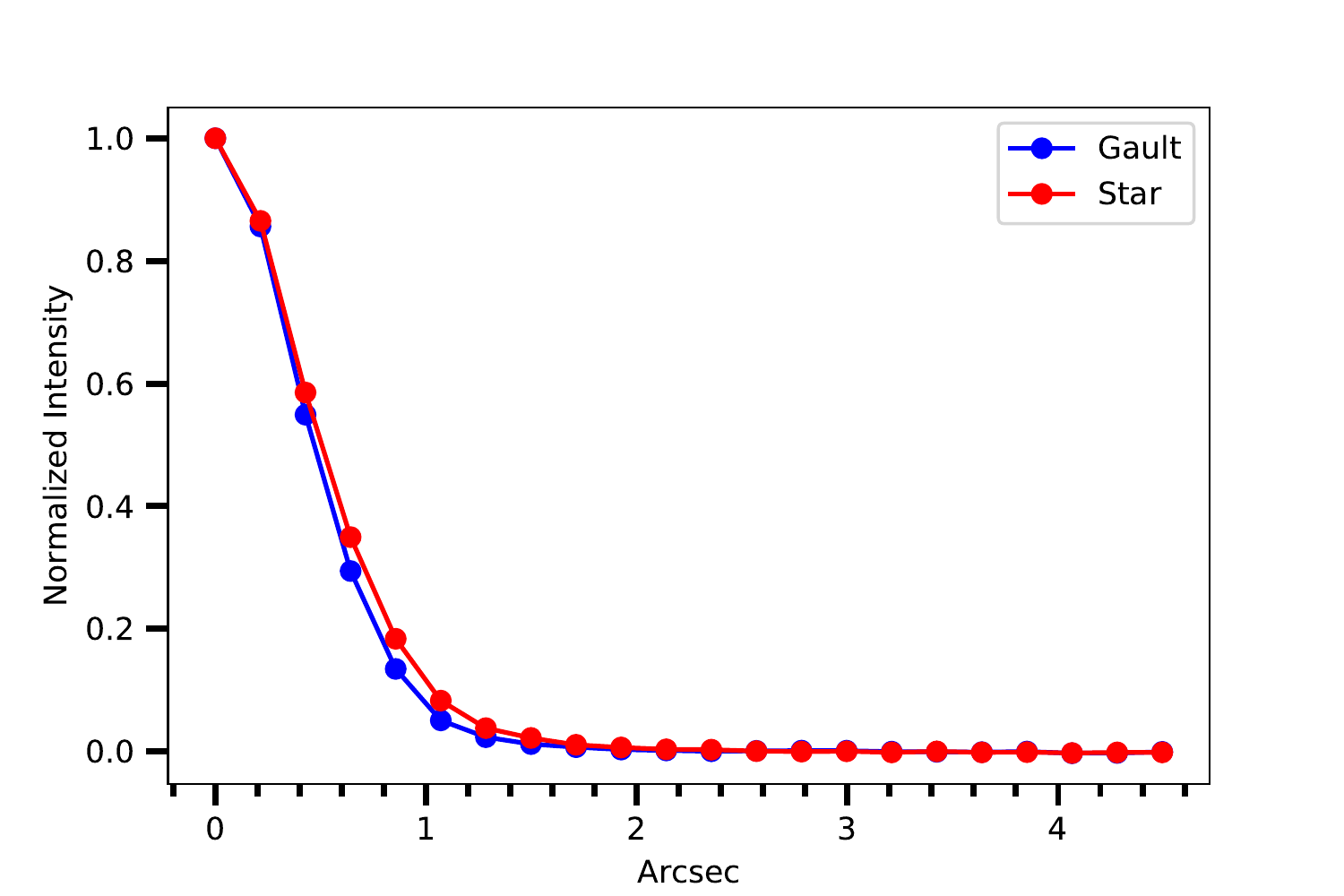}
\caption{Gault's profile from UT 2020 Aug 27, measured with the ALFOSC camera on the NOT telescope.  Also shown in the plot is the profile of standard star Feige 11.  Pixel scale: 0.214\arcsec / pixel.
\label{fig:profileNOT}}
\end{figure}

As another way to determine whether our photometry might be contaminated by dust emission,  we measured Gault's radial profile using both NOT and HST images.  Figure \ref{fig:profileNOT} shows Gault's profile extracted from one of the NOT images, compared with the profile from the standard star Feige 11 (\cite{landolt2009}) imaged using sidereal tracking rates.  For both objects, the profiles were extracted the same way and have been normalized to the peak intensity.  The two profiles have essentially the same Full Width at Half Max (FWHM) of 4.6 pixels, or 1.0\arcsec; the very slight discrepancy between the two profiles can be attributed to the different seeing of the two images.  This result suggests that we observed the bare nucleus of Gault.  \\

\begin{figure}
\plotone{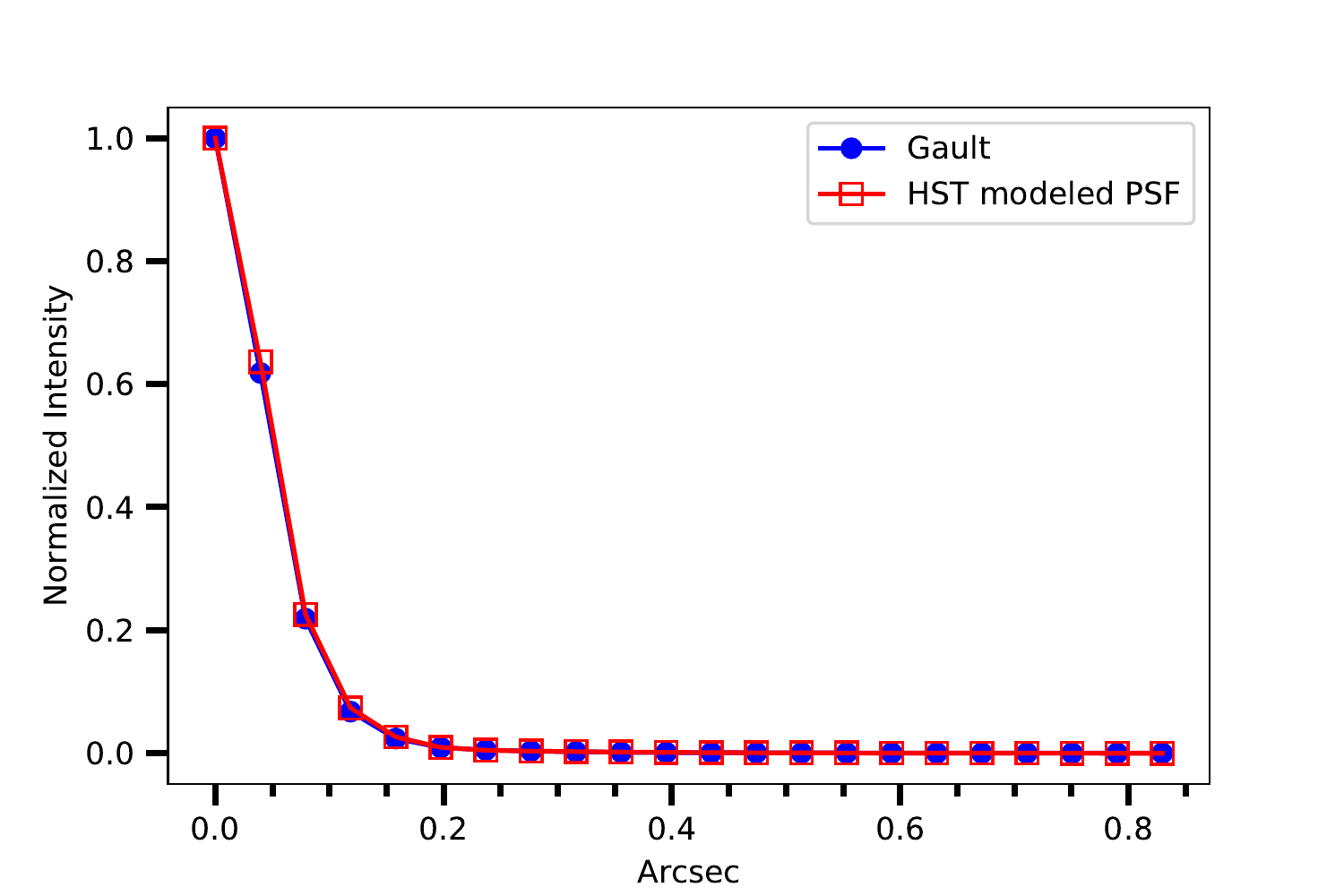}
\caption{Gault's profile from from UT 2020 Aug 26, as measured with the WFC3 camera on the Hubble Space Telescope.  Also shown in the plot is a synthetic PSF of WFC3.  The two profiles are identical.  Pixel scale: 0.0395\arcsec / pixel. 
\label{fig:profileHST}}
\end{figure}

\begin{figure}
\plotone{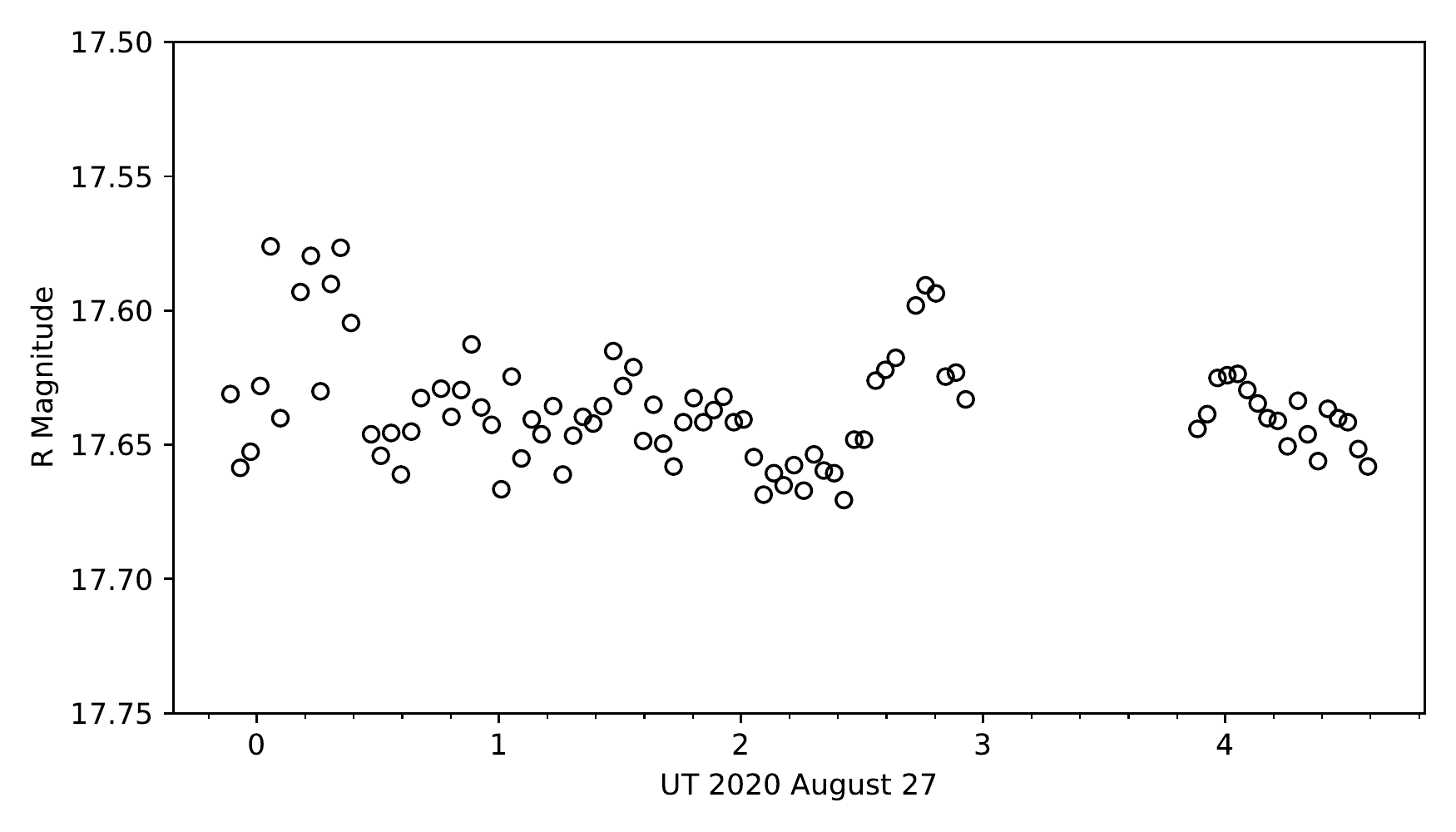}
\caption{R magnitude of Gault as a function of time, taken on UT 2020 Aug 27, at the NOT telescope.
\label{fig:lightcurve}}
\end{figure}

The HST images provided a higher resolution measure of  Gault's profile.  In order to generate an HST point spread function (PSF) for comparison with the Gault profile, we used the TinyTIM software (\cite{krist2011}), version 7.5.  A  PSF for the WFC3 camera and F350 LP filter combination was generated, then scaled to the same pixel scale as the Gault images.  Figure \ref{fig:profileHST} shows Gault's profile extracted from one of the HST images, compared with the model PSF.   The two profiles are essentially identical, providing another confirmation that we were observing Gault's nucleus.  \\

To set a limit to the presence of coma we used the relation between the surface brightness, $\Sigma(\theta)$ measured at angular radius, $\theta$, and $m_C$,  the  apparent magnitude of the coma within a circle of radius $\theta$, from \cite{jewitt1984},

\begin{equation}
m_C = -2.5\log_{10}(2\pi \theta^2) + \Sigma(\theta).
\label{JD}
\end{equation}

\noindent  This relation is based on the assumption that $\Sigma(\theta) \propto \theta^{-1}$, as expected for a steady-state flow, and as is commonly observed in the inner comae of comets before the effects of radiation pressure deflect the ejected particles from their initial paths.   \\

From the HST surface brightness profile, we set a limit to $\Sigma(0.2\arcsec) \ge$ 25.0 magnitudes arcsecond$^{-2}$.  Substituting into Equation (\ref{JD}) gives $m_C \ge$ 23.5, at a time when the red magnitude of Gault was $m_R$ = 17.64.   We conclude that any contribution from a steady-state near-nucleus coma is fainter than the integrated magnitude by $|m_C - m_R|$ = 5.9 magnitudes.  Such a coma could contribute no more than $10^{(-0.4 |m_C - m_R|))}$ = 0.5\%  to the measured signal and thus cannot account for the ten-times larger photometric variations observed in Gault.  We cannot, however, use the surface brightness profile to constrain the presence of dust particles packed closely to the nucleus, for example sub-orbital particles moving in temporarily bound orbits close to the nucleus. \\

We use the limit to the coma obtained from the surface brightness profile to set a limit to the production of dust.  The cross-section of Gault is $C_n = \pi r_n^2$, where $r_n$ = 2 km is the estimated radius (\cite{sanchez2019}).  The dust cross section, $C_d$, inside a circle having radius $\theta_i$ = 0.2\arcsec~is roughly $C_d < C_n 10^{(-0.4 |m_C - m_R|)}$, or $C_d < 63,000$ m$^2$.  We suppose that the dust has average radius $\bar{a}$ and  is moving radially outwards at speed $V$.  The time taken for dust to travel angular distance $\theta$ is $\tau \sim \theta \Delta/V$, with $\theta$ expressed in radians, and the mass supply and loss rates needed to maintain steady state are $\dot{M} \sim 4\rho \bar{a} C_d/(3\tau)$, or

\begin{equation}
\frac{dM}{dt} = \frac{4\rho \bar{a} C_d V}{3 \theta \Delta}
\end{equation}

\noindent  We set   $\bar{a}$ = 200 $\mu$m and $V$ = 0.15 m s$^{-1}$, respectively (\cite{jewitt2019a}), $\Delta$ = 1.32 AU (Table 1) and assume $\rho$ = 1700 kg m$^{-3}$ (see below), to find $\dot{M} \lesssim 0.02$ kg s$^{-1}$.  This compares with peak mass loss rates $\dot{M} =$ (20 to 40) kg s$^{-1}$ during the maximum tail formation phase (\cite{jewitt2019a}).

%
%

\subsection{Lightcurve}

Except for small amplitude objects, most asteroid lightcurves show two clear maxima and minima over one rotation period, caused by the changing cross section of an irregularly shaped body.  In contrast, Gault's brightness variations show multiple (four or five) small peaks, all with very small ($\le 0.025 -  0.05$ mag) amplitudes (Figure \ref{fig:lightcurve}).  These small, short-term brightness variations could be due to albedo variations, to local topographical deviations from symmetry (i.e.~"lumps'' on the surface), or to a combination of both. \\

\begin{figure}
\plotone{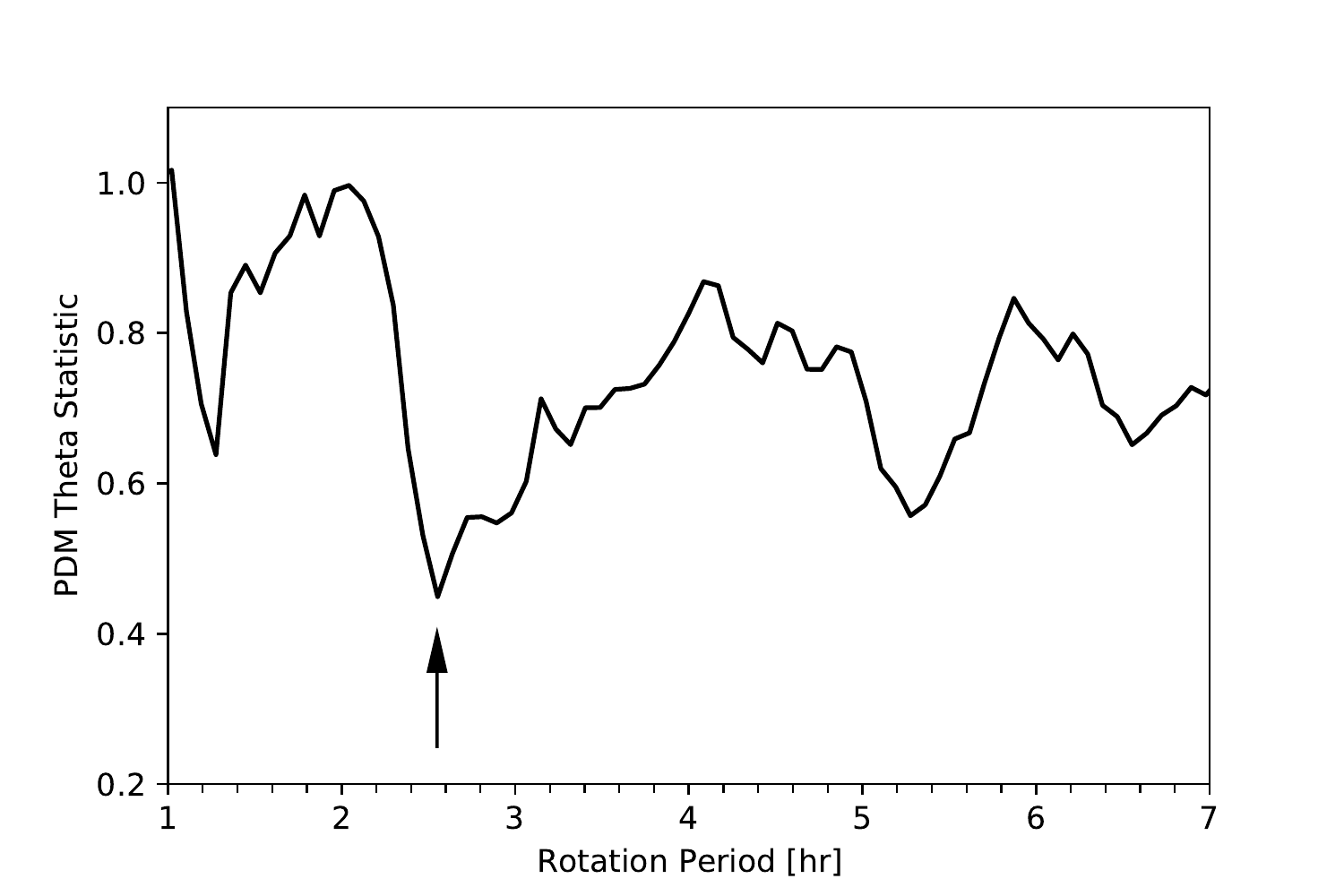}
\caption{Result from PDM algorithm, showing that the most likely rotation period from Gault's lightcurve is 2.55 hr (indicated by arrow).  A slightly shallower minimum near $\sim 5.3$ hr is most likely due to subharmonics.
\label{fig:PDM}}
\end{figure}

\begin{figure}
\plotone{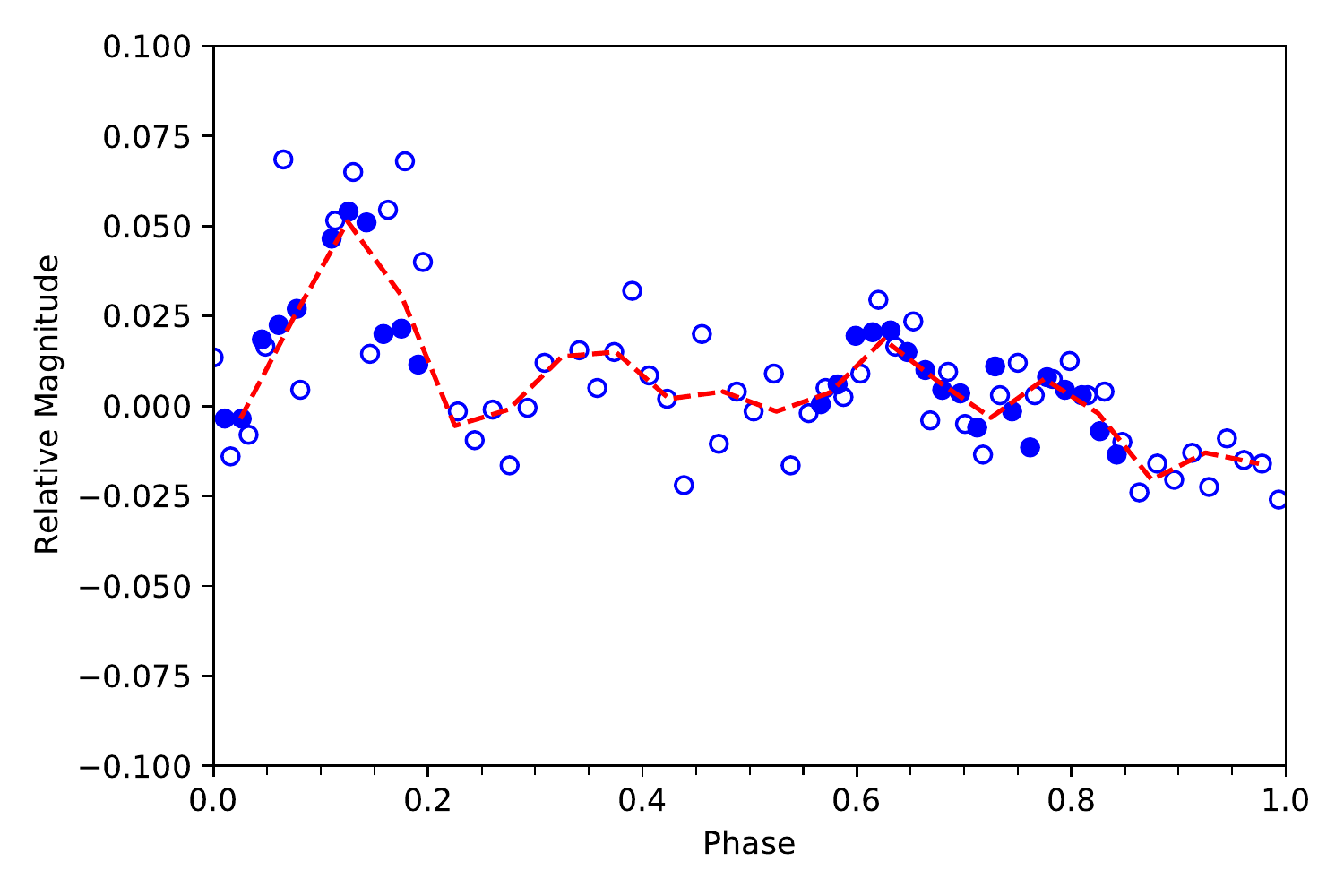}
\caption{Phase plot from Gault's lightcurve, based on a 2.55 hr rotation period.  The phased data are divided into 20 phase bins, and the dashed line is the median of the data in each phase bin.   The data suggests $\sim 4 - 5$ peaks,  and the very small lightcurve amplitudes are consistent with a spinning top shape.
\label{fig:phaseplot}}
\end{figure}

To look for periodicity in the brightness variations, we use the phase dispersion minimization (PDM) algorithm (\cite{stellingwerf2011}.  The correct period would result in a minimum theta statistic, as defined in \cite{stellingwerf1978}.  The result of the PDM algorithm is shown in Figure \ref{fig:PDM}, and shows a minimum at 2.55 hr, with a shallower minimum near $\sim 5.3$ hr.  For reasons explained below, we believe the 2.55 hr is the correct rotation period, with the 5.3 hr most likely due to subharmonics.   To verify the 2.55 hr rotation period, Figure \ref{fig:phaseplot} shows the phase plot for this period; the phased data, which cover nearly 2 full rotations, show good overlap over the full rotational phase.  To guide the eye we also plot the median of the data divided into 20 phase bins.  This median line suggests that Gault has four or five peaks with amplitudes of a few $\times$ 0.01 magnitude.   Periods outside the range 2.45 hr and 2.65 hr fail to generate convincing phase plots.
\\

As noted above, several groups reported no detectable rotation period for Gault during the 2019 apparition (\cite{moreno2019}, \cite{sanchez2019}, and \cite{lin2020}), but  we believe that this was due to the dust emission that existed throughout 2019.  The dust velocity was measured at $\sim 0.15$ m s$^{-1}$ (\cite{jewitt2019a}); at this velocity, assuming geocentric distance $\Delta  = 1.4$ AU (as in \cite{sanchez2019}, it would take $\sim$8 months for the dust to traverse a 3-arcsec radius photometry aperture.  The dust contribution would  obscure any rotational modulation due to the nucleus, leaving the slight modulation of the brightness due to nucleus rotation undetected in the 2019 observations.   We note that our result is longer than the 2-hr rotation period reported by \cite{kleyna2019}, and  inconsistent with those  of \cite{ferrin2019} and \cite{carbognani2020}, who reported a $\sim$3.36 hr rotation period.   \\

The 2.55 hr rotation period is near the empirical $P \sim$ 2.2 hr critical period for asteroids (\cite{pravec2000}), and suggests rotational instability as the simplest explanation for Gault's mass loss.   The critical period for rotational instability of a spherical fluid (i.e.~strengthless) body in rotation at period $P$ occurs for density

\begin{equation}
\rho_c = \frac{3\pi}{G P^2}
\label{crit}
\end{equation}

\noindent where $G = 6.67\times10^{-11}$ N kg$^{-2}$ m$^2$ is the gravitational constant.  Substituting $P$ = 2.55 hour gives $\rho_c$ = 1700 kg m$^{-3}$ which is comparable to, but slightly higher than, the densities of (101955) Bennu, $\rho = 1190\pm13$ kg m$^{-3}$ (\cite{scheeres2019}) and of (162173) Ryugu, $\rho = 1190\pm20$ kg m$^{-3}$  (\cite{watanabe2019}).   The higher density, taken at value, would imply that Gault is less porous than either of these two asteroids, perhaps consistent with its larger size (Bennu and Ryugu are, respectively, about 0.5 km and 1 km in diameter, compared with Gault at $\sim$4 km).  However, density estimates using Equation (\ref{crit}) are crude, since Gault is unlikely to be either perfectly spherical or strengthless, and we do not wish to over-interpret the result.  \\

 
%
%

\section{Discussion}

\subsection{Shape}
  

The small lightcurve amplitude of Gault may result from a spin axis pointing nearly along the line of sight, or from a nearly symmetric shape.  We cannot rule out the former hypothesis, but we prefer the latter for the following reasons.  We notice that Gault shares several similarities with asteroid (101955) Bennu, the target of NASA's OSIRIS-REx mission (see Table 2).  Both asteroids display lightcurves with very small amplitudes:  0.025 - 0.05 mag for Gault, 0.03 - 0.06 mag for Bennu (\cite{hergenrother2019}).  (We note that there is an earlier version of Bennu's lightcurve  with larger amplitudes (\cite{hergenrother2013}), but here we refer to the lightcurve measured by OSIRIS-REx, since the 4 - 18 deg phase angles of these measurements are much more comparable to ours).  Both asteroids lose mass episodically, although at $\sim 10^{-7}$ kg s$^{-1}$ (\cite{lauretta2019a}), Bennu's mass loss rate is 8 orders of magnitude smaller than Gault's 40 kg s$^{-1}$ at its peak (\cite{jewitt2019a}).  Both asteroids eject large and slow moving particles: 1-10 cm diameter and 0.07 - 1 m s$^{-1}$ for Bennu (\cite{lauretta2019a}), $\sim 0.4$ mm diameter and 0.15 m s$^{-1}$ for Gault (\cite{jewitt2019a}, \cite{kleyna2019}).
Images of Bennu reveal a rubble pile asteroid with a top-like shape (\cite{lauretta2019b}).  The top shape is also shared by (162173) Ryugu, although mass loss has not been reported for this object (\cite{watanabe2019}).  Given the preponderance of the top shape among small asteroids (e.g., \cite{nolan2013}), and Gault's similarities with Bennu, we suspect Gault is also top-shaped.  \\

The top-like shape has been seen in several small asteroids, such as 1994 KW4 Alpha (the primary of a binary system, \cite{ostro2006}),  (65803) Didymos (\cite{pravec2006}), 2008 EV5 (\cite{busch2011}), and most recently in Bennu and Ryugu.  Several hypotheses have been proposed for their shape.  Small asteroids can be spun up or down by the YORP effect on timescales $10^5 - 10^6$ yrs (\cite{rubincam2000}), so that over the lifetime of the asteroid, material from higher latitudes is driven toward the equator by centrifugal forces (\cite{walsh2012}).  Rubble pile asteroids like Bennu and Ryugu are believed to have formed from the reaccumulation of fragments produced in a catastrophic asteroid collision (\cite{michel2001}, \cite{michel2020}). Support for an early origin comes from images showing  craters imprinted on the equatorial ridge at both Bennu and Ryugu (\cite{walsh2019}, \cite{hirata2020}).   Alternatively, fast spinning rubble pile asteroids are prone to deformation, both internally and at the surface, and may have acquired the top shape that way (\cite{hirabayashi2020}).   Bennu, Ryugu and Gault thus could have acquired their shape at the time of formation, or soon after.   \\

\subsection{Surface}

The surfaces of Bennu and Ryugu provide a glimpse of what Gault's surface might look like.  Bennu's surface is much rougher than expected, with hundreds of 10m-size boulders and even more at the 1-m scale (\cite{lauretta2019b}).  Fine grains exist, but in limited amounts.   
Gault dust tail measurements also suggest a surface lacking in small particles (\cite{jewitt2019a}), and the same was also reported for (3200) Phaethon (\cite{ito2018}).   Small particles are apparently scarce on small rubble pile asteroids, whether they are active or not; the reason for the scarcity of small particles is still unclear.\\

\subsection{Mass shedding threshold}

The level of mass shedding in fast spinning rubble pile asteroids is a competition between centrifugal force and gravity, complicated by the effects of inter-particle friction and cohesion.   To investigate the different regimes of rotational mass shedding, we calculate the Froude number $Fr$ for those active asteroids suspected to be losing mass due to rotational instability.  For example, $Fr$  is often used to investigate granular flow regimes, such as in rotating drums (\cite{mellmann2001}).  The Froude number is defined as the ratio between the centrifugal and gravitational forces, i. e., $Fr = \omega^2 R/g$, where $\omega$ is the angular rotation speed of the granular system, $R$ its radius and $g$ the gravitational acceleration.   $Fr$ is proportional to the square of rotation speed, reflecting the fact that the rotation speed is the most important factor in controlling the flow of the particles.  




Table 3 lists the active asteroids for which diameter, rotation period, mass loss rate data are available; it is a subset of Table 2 of Jewitt et al. (2015).   The density is available for most; we indicate where the density is assumed.    In Figure 7 we plot the normalized mass loss rate, $\left( dM/dt \right) / (4 \pi R^2$), as a function of the Froude number.   We note the following:

\begin{enumerate}

\item For fast spinning rubble pile asteroids, the onset of mass shedding starts at $Fr \sim 0.5$.  Particles can be lost at $Fr < 1$  because particle escape is aided by other factors not included in the Froude number, such as surface slopes, and radiation pressure forces (McMahon et al. 2020).  

\item Asteroids Bennu, (3200) Phaethon, and 133P/(7968) Elst-Pizarro all have $Fr \sim 0.5$ but Bennu's mass loss rate is 6 orders of magnitude smaller than the others'.  The large gap between Bennu and the others is no doubt due to the different sensitivity limits of ground-based vs. spacecraft measurements, and as more data become available, we expect the gap to be filled in.  The clustering of $Fr \sim 0.5$ indicates that once mass shedding starts at $Fr \sim 0.5$, a single $Fr$ could be associated with a wide range of mass loss rates.  
  
\item  Beyond the $Fr \sim 0.5$ threshold, mass loss generally increases with $Fr$, consistent with rotation being the main driver of mass loss.   As $Fr$ grows, there must be a critical $Fr$ beyond which the asteroid is completely blown apart by centrifugal forces.  This is most likely what happened to asteroid 2013 R3 (\cite{jewitt2014b}, \cite{hirabayashi2014}), but there is no measurement of 2013 R3's rotation rate, so it was not included on this Figure. 
Hopefully future observations will reveal more insight into this extreme rotation regime. 
\end{enumerate}

Finally, Figure 7 suggests that rotational instability can contribute to mass loss in some asteroids where the cause of activity has been ambiguous.  For example, activity in asteroid (3200) Phaethon has been tentatively attributed to thermal fracture and/or desiccation cracking, while thermal fracturing, volatile release by dehydrated rocks, and meteoroid impacts have been postulated for Bennu.  While the extreme temperature cycling in Phaethon may contribute dust ejection, Figure 7 suggests that rotation may also play a role.  \cite{lauretta2019a} rejected rotational disruption as the origin of Bennu's ejected particles due to the fact that the particles were in retrograde orbits.  \cite{mcmahon2020} pointed out that the inclusion of non-Keplerian forces like solar tides and radiation pressure could change the orbital elements significantly over a timescale of $\sim100$ days.   Such fast evolution of the orbits could explain the retrograde orbits of Bennu's dust particles, allowing rotational instability to be reconsidered as a possible cause of dust ejection.  

\clearpage

\section{Conclusions}

Observations of Gault in August 2020 show both a point-like PSF and a faint absolute magnitude consistent with the absence of near-nucleus dust.  We find that

\begin{enumerate}

\item Gault shows a rotational lightcurve with period $2.55 \pm 0.10$ hr. This  corresponds to the critical period for rotational instability of a strengthless sphere having density $\rho$ = 1700 kg m$^{-3}$. 

\item The largest feature in the lightcurve  has amplitude $\sim$5\%, showing that the body when projected into the plane of the sky is closely symmetric.  

\item The overall properties of Gault, including the density, the small lightcurve amplitude, and the low velocity ejection of large particles, are strongly reminiscent of asteroid (101955) Bennu.  Like Bennu, Gault is probably a rubble pile asteroid shaped like a top.

\item  The Froude number $Fr$, defined here as the ratio between centrifugal force at the equator and gravitational force, could be a useful predictor of the onset of mass loss in fast spinning rubble asteroids.  Observational data indicate that mass loss starts at $Fr \sim 0.5$, and once this threshold is reached, there is a large spread of activity level.  The Froude number is useful for identifying those asteroids whose activity can be adequately explained by rotational instability, without the need for other less common mechanisms. 
\end{enumerate}

\clearpage

\section*{Note}
As we were about to submit this manuscript, we learned of a preprint by Purdum et al. (http://ui.adsabs.harvard.edu/abs/2021arXiv210213017P/abstract) that also presented a rotational lightcurve for Gault.  The preprint's results are consistent with ours.

\acknowledgments

We thank the NOT staff (particularly Teet Kuutma) and HST staff for helping with the observations.  This research is based on observations made with the NASA/ESA Hubble Space Telescope obtained from the Space Telescope Science Institute. These observations are associated with program GO 15972. Y. K. and J. A. acknowledge funding by the Volkswagen Foundation.  J.A.'s contribution was made in the framework of ERC Starting Grant No 757390.

\bibliography{Gault2021_final}{}
\bibliographystyle{aasjournal}

\clearpage

\begin{deluxetable*}{ccccccc}
\tablenum{1}
\tablecaption{Observations\label{tab:observations}}
\tablewidth{0pt}
\tablehead{
\colhead{Telescope} & \colhead{UT Date} & \colhead{Exp$^{ a}$} & \colhead{Filter} &
 \colhead{$r_H^{ b}$} & \colhead{$\Delta ^{ c}$} & \colhead{$\alpha ^{ d}$} 
}
\decimalcolnumbers
\startdata
NOT & 2020 Aug 26  & $88 \times 200$ s & R & 2.21 & 1.34 & 17.2 \\
HST WFC3 & 2020 Aug 28 & $4 \times 245$ s & F350LP & 2.21 & 1.32 & 15.9 \\
\enddata

\noindent $^a$  Image exposure time, seconds \\
\noindent $^b$  Heliocentric distance, AU \\
\noindent $^c$  Geocentric distance, AU \\
\noindent $^d$  Phase angle, degree \\

\end{deluxetable*}

\clearpage 

\begin{deluxetable*}{cccccccl}
\tablenum{2}
\rotate
\tablecaption{Comparison with Asteroid (101955) Bennu \label{tab:comparison}}
\tablewidth{0pt}
\tablehead{
\colhead{Asteroid} & \colhead{$D^{ a}$}  & \colhead{$P^{ b}$} & \colhead{Shape} & \colhead{$\Delta m^{ c}$} 
& \colhead{Ejected particle size} & \colhead{Ejected particle velocity} & \colhead{References}
}
\decimalcolnumbers
\startdata
(6478) Gault & 4 & 2.55 &  & 0.03 - 0.07 & $\sim 400 \mu$m & $\sim 1.4$ m s$^{-1}$ & \cite{jewitt2019a}, this work \\
(101955) Bennu & 0.506 & 4.29  & Top-like & 0.03 - 0.06 $^{ 1}$  & $\sim 1$ cm & $0.07 - 3$ m s$^{-1}$ & \cite{hergenrother2019} \\
\enddata

\noindent $^a$ Diameter [km] \\
\noindent $^b$  Rotation period [hr] \\
\noindent $^c$  Lightcurve amplitude [mag] \\

\tablecomments{
\noindent $^{ 1}$  The lightcurves of Bennu have been measured from both the ground (\cite{hergenrother2013}) and with spacecraft (\cite{hergenrother2019}).  Here we refer to the latter because the observations were made at small phase angles (4 - 18 deg) that are more comparable to ours.}

\end{deluxetable*}

\clearpage

\begin{deluxetable*}{lcccccl}
\tablenum{3} 
\rotate
\tablecaption{Fast Spinning Active Asteroids \label{tab:Froude}}
\tablewidth{0pt}
\tablehead{
\colhead{Asteroid} & \colhead{$D^{ a}$} & \colhead{$P^{ b}$} & \colhead{$\rho^{ c}$} & \colhead{$dM/dt^{ d}$}  & \colhead{$Fr^{ e}$}  
& \colhead{References} 
}
\startdata
(6478) Gault & 4 & 2.55  & 1190$^*$ & $\sim 35$ & 1.4 & \cite{jewitt2019a}, \cite{sanchez2019} \\
(101955) Bennu & 0.506 & 4.29 & 1190 & $10^{-7}$ & 0.5 & \cite{lauretta2019a} , \cite{lauretta2019b}, \cite{hergenrother2019} \\
133P/(7968) Elst-Pizarro & 3.9  & 3.47 & 1300$^*$ &  1.6 & 0.7  &  \cite{hsieh2004}, \cite{jewitt2014a}\\
(3200) Phaethon & 5.1  & 3.6 & 1670 &  3  & 0.5  & \cite{jewitt2013}, \cite{hanus2016}, \cite{hanus2018}\\
331P/Gibbs & 1.8  & 3.24 &  1000$^*$ &  25  & 1.0 & \cite{stevenson2012}, \cite{drahus2015} \\
P/2017 S5 (ATLAS) & 0.9 & 2.88 & 1000 & 5 & 1.3 & \cite{jewitt2019b}\\
332P/Ikeya-Murakami & 0.55 & 2.2$^*$ & 1000$^*$ & 257 $^f$ & 2.25 & \cite{jewitt2016}\\
\enddata
\noindent $^a$ Diameter [km] \\
\noindent $^b$  Rotation period [hr] \\
\noindent $^c$  Density [kg/m$^3$] \\
\noindent $^d$  Dust mass loss rate [kg s$^{-1}$]  \\
\noindent $^e$  Froude number  \\
\noindent $^*$  Assumed \\
\noindent $^f$ The mass loss rate is estimated as the total ejected mass $2 \times 10^9$ kg spread out over 3 months of activity. \\

\end{deluxetable*}

\clearpage

\clearpage

\clearpage

\clearpage

\clearpage

\clearpage

\clearpage

\begin{figure}
\plotone{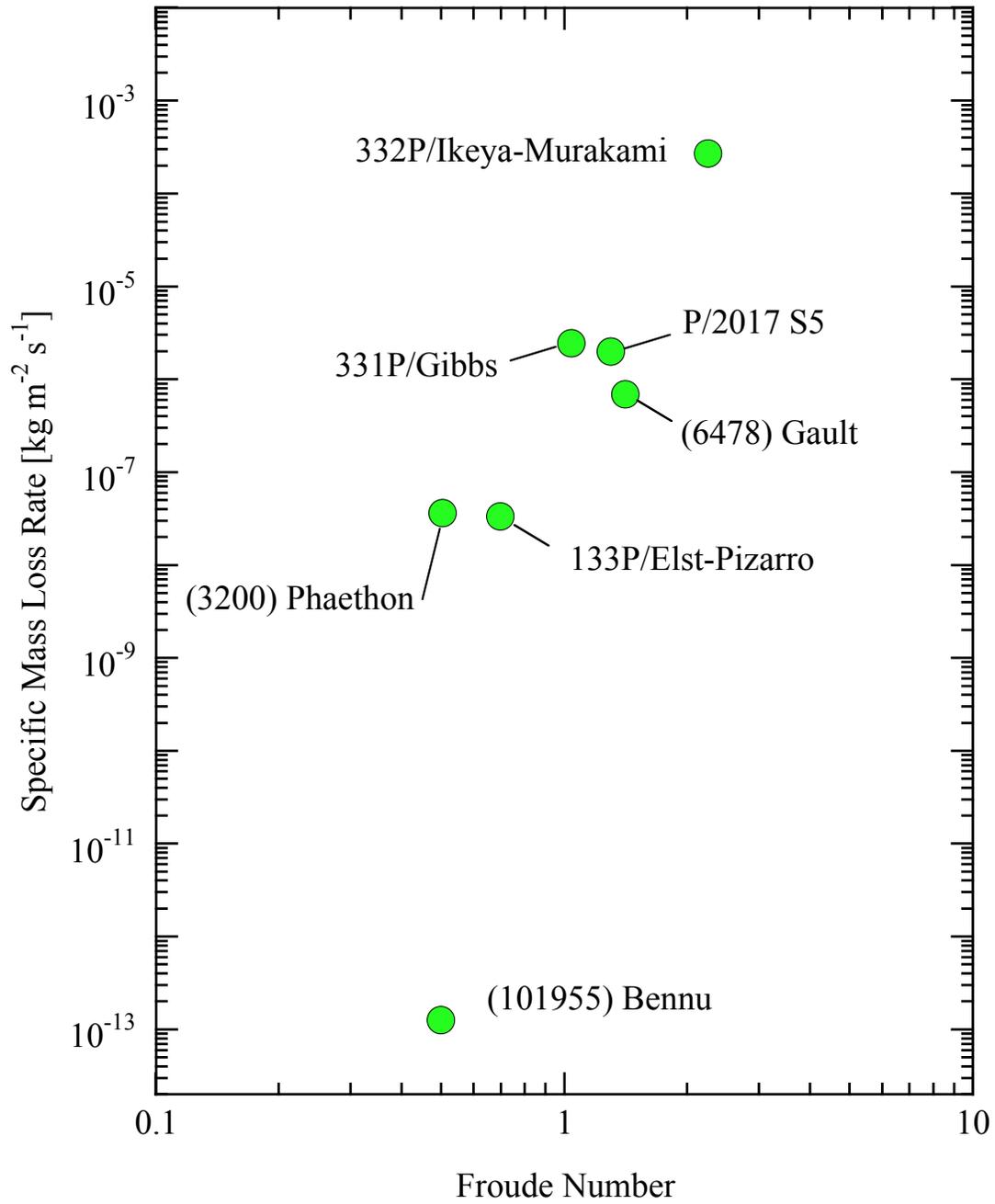}
\caption{Normalized mass loss rate (mass loss rate divided by surface area) as a function of Froude number.
\label{fig:Froudeplot}}
\end{figure}



\end{document}